\documentclass[prd,aps,11pt,eqsecnum,
nofootinbib,tightenlines,superscriptaddress, preprintnumbers]{revtex4}

\def\lsim{\raise0.3ex\hbox{$<$\kern-0.75em\raise-1.1ex\hbox{$\sim$}}}
\def\gsim{\raise0.3ex\hbox{$>$\kern-0.75em\raise-1.1ex\hbox{$\sim$}}}

\usepackage{graphicx}

\begin{document}

\title{Jet hadrochemistry as a characteristics of jet quenching}

\author{Sebastian Sapeta}
\affiliation{Department of Physics, CERN, Theory Division, CH-1211 Geneva 23, Switzerland}
\affiliation{M. Smoluchowski Institute of Physics, Jagellonian University,
   Reymonta 4, 30-059 Cracow, Poland}

\author{Urs Achim Wiedemann}
\affiliation{Department of Physics, CERN, Theory Division, CH-1211 Geneva 23, Switzerland}

\preprint{CERN-PH-TH/2007-111}

\begin{abstract}
Jets produced in nucleus-nucleus collisions at the LHC are expected to be strongly modified 
due to the interaction of the parton shower with the dense QCD matter. Here, 
we point out that jet quenching can leave signatures not only in the longitudinal and transverse 
jet energy and multiplicity distributions, but also in the hadrochemical composition of 
the jet fragments. In particular, we show that even in the absence of medium
effects at or after hadronization, the medium-modification of the parton shower can result in
significant changes in jet hadrochemistry. 
We discuss how jet hadrochemistry can be studied within the high-multiplicity
environment of nucleus-nucleus collisions at the LHC.
\end{abstract}

\maketitle

\section{Introduction}
\label{sec1}

Highly energetic partons, propagating through dense QCD matter, are fragile. Their
fragmentation pattern changes in the presence of a strongly interacting medium. This is
reflected in the medium-modification of their hadronic remnants  in
nucleus-nucleus collisions at RHIC~\cite{Adcox:2004mh,Back:2004je,Arsene:2004fa,Adams:2005dq}
and will soon be further explored at the higher LHC collider energies~\cite{Carminati:2004fp,Alessandro:2006yt,D'Enterria:2007xr,Takai:2004nm}. The generic suppression of
high-$p_T$ single inclusive hadron spectra~\cite{Adcox:2001jp,Adler:2003au,Adler:2002xw,Adams:2003kv,Back:2003qr,Arsene:2003yk,Adams:2004wz}
 and triggered particle correlations~\cite{Adler:2002tq,Magestro:2005vm,Adams:2006yt}
in \mbox{A-A} collisions 
at RHIC support this picture. These measurements can be largely accounted for in models of 
radiative parton energy loss~\cite{Wang:2003aw,Dainese:2004te,Eskola:2004cr,Gyulassy:2000gk,Hirano:2002sc,Renk:2006sx}, which assume
that additional medium-induced parton splitting leads to a softening of the parton shower,
and thus to an energy degradation of the leading hadrons. If the leading hadronic fragment 
in a jet carries less energy, then - due to energy-momentum conservation - either the subleading 
fragments carry more, or their multiplicity increases accordingly.
The resulting changes expected for the longitudinal~\cite{Salgado:2003rv,Pal:2003zf,Borghini:2005em,Majumder:2004pt}
  and transverse~\cite{Salgado:2003rv,Armesto:2004pt,Vitev:2005yg,Polosa:2006hb,Lokhtin:2006rj} 
jet energy and jet multiplicity distributions have been discussed in the context of models, 
which account successfully
for the suppression of single inclusive hadron spectra. In this paper, we study a third class
of jet characteristics, which may show significant medium effects: measurements of the 
hadrochemical composition of jet fragments.

\begin{figure}[t]
\includegraphics[scale=.375,angle=0]{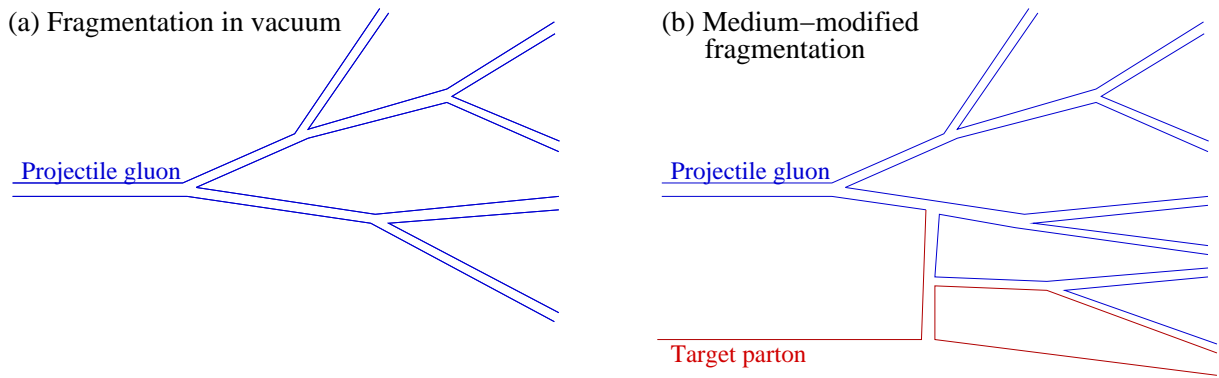}
\caption{Sketch of an entirely gluonic parton shower in the large $N_c$ limit, where gluons
are represented as pairs of $q\bar{q}$ fermion lines, and quarks as single lines.
(a) Fragmentation of the gluon in the vacuum. (b) Interaction of the gluon with a target
quark in the medium via a single gluon exchange. This interaction changes the color flow
and may affect hadronization, see text. }
\label{fig1}
\vskip-0.05in
\end{figure}

There are several reasons of why one expects parton energy loss to affect jet hadrochemistry.
In particular, in all models of radiative parton energy loss
\cite{Gyulassy:1993hr,Baier:1996sk,Zakharov:1997uu,Wiedemann:2000za,Gyulassy:2000er,Wang:2001if}, the interaction of a parent parton 
with the QCD medium transfers 
color between partonic projectile and target. This changes the color flow in the parton shower 
and is thus likely to affect hadronization. To illustrate this point, we have sketched in Fig.~\ref{fig1}a
an entirely gluonic parton shower in the large-$N_c$ approximation, where gluons are 
represented as $q\bar{q}$-dipoles. At large $N_c$, the color singlet prehadronic subsystems
at the end of the perturbative evolution of the shower may be identified with the connected
fermion lines in this diagram. Fig.~\ref{fig1}b illustrates the conceivable effects of a single gluon 
exchange between the projectile gluon and a target quark in the medium. The multiplicity in the
shower increases and the momentum distribution may soften and widen accordingly. Also,
in general, the color singlet prehadronic subsystems at the end of the perturbative evolution 
will have a different distribution in invariant mass. This makes it likely that their subsequent
fragmentation into hadrons results in a hadrochemical distribution different from that of a 
gluon fragmenting in the vacuum. This model-dependent example illustrates that existing
models of parton energy loss, though formulated in an entirely partonic language, contain
one possible ingredient for a medium-induced modification of the hadrochemical composition
of jets, namely: color exchange between projectile and target.

In addition, flavor or baryon number could be exchanged between medium and projectile. 
For instance, subleading partons in the jet may be subject to hadronization 
via recombination with partons from the medium~\cite{Fries:2003vb,Molnar:2003ff,Greco:2003mm,Hwa:2002tu,Fries:2003kq,Maiani:2006ia,Fries:2003fr}, a mechanism which is conjectured to 
underlie baryon enhancement of identified single inclusive spectra at intermediate $p_T$.
If parton energy loss involves non-negligible recoil effects (a.k.a. collisional energy 
loss~\cite{Braaten:1991jj,Djordjevic:2006tw,Adil:2006ei}), then
one may also speculate that the hard parton kicks components of the medium into the jet cone.
Depending on the quantum numbers carried by the components kicked, this will affect
jet hadrochemistry. In short, medium-induced modifications of
jet hadrochemistry are a direct (for the case of color transfer) or conceivable (for the case of recombination, flavor and baryon number transfer) consequences of the current models of 
jet quenching. They provide complementary information about the microscopic mechanism
underlying parton energy loss. Thus, jet hadrochemistry may be a valuable tool for
establishing the properties of the matter produced in heavy ion collisions. 

The interfacing of a parton shower with the hadronic world is a significant modeling task with very
little guidance from theory, even for the simplest systems without medium-modifications,
such as jets in $e^+e^-$ collisions. This task is more complex for the case of a medium-modified 
jet, since it depends now additionally on information about the exchange of momentum, color, 
flavor and baryon number between the partonic projectile and the medium. In the present work, 
we shall largely bypass these complications. We merely observe that even if the exchange of 
quantum numbers between projectile and target is disregarded, a change of hadrochemical 
composition can be expected for the generic case that the distribution of partons in
invariant mass is modified. It is a model of this type \cite{Borghini:2005em}, which we introduce in section~\ref{sec2}.
Since this model does not implement any of the effects mentioned above, it may be expected
to underestimate the medium-modifications of jet hadrochemistry. Specific signatures of 
hadrochemical modifications, indicative e.g. of color, flavor or baryon number transfer, may
be established as deviations from this model. 
In section~\ref{sec3}, we discuss the hadrochemical composition of the high-multiplicity 
soft background, in which medium-modified jets are 
immersed in heavy ion collisions. We emphasize that the hadrochemical composition of
this background differs significantly from that of a proton-proton collision, and it also differs
form the vacuum fragmentation of a jet. This difference in composition will help to characterize
jet hadrochemistry in nucleus-nucleus collisions. Our main conclusions are summarized
in section~\ref{sec4}.

\section{Jet multiplicity distributions and their medium-modification}
\label{sec2}

In this section, we introduce and study a specific model of how the single inclusive hadron 
distribution in a jet is composed of different hadron species, and how it 
changes due to medium-effects. 

\subsection{Single inclusive intrajet distributions in the absence of medium modification}
\label{sec2a}

The single inclusive distribution $dN^h/d\xi$ of hadrons inside a jet, plotted versus the
logarithm of the hadronic momentum fraction $\xi = \ln\left[ 1/x \right]$, $x=p_h/E_{\rm jet}$,
shows a characteristic hump-backed plateau. Measurements of this distribution in
$e^+e^-$ collisions and hadronic collisions are well described in the MLLA (modified
leading logarithmic approximation) formalism~\cite{dokshitzer:lnpi,Dokshitzer:1988bq,Dokshitzer:1991fc,Fong:1990nt}, supplemented by local parton-hadron
duality (LPHD) \cite{Dokshitzer:1991fc,Azimov:1984np,Azimov:1985by}. Here, the MLLA is a perturbative calculation of the parton distribution
$D_{q,g}(\xi, \tau, \lambda)$ inside a quark or gluon jet, which achieves double and single logarithmic accuracy in 
$\xi = \ln\left[ 1/x \right]$ and $\tau = \ln\left[ Q/\Lambda\right]$ where $\lambda = \ln\left[ Q_0/\Lambda\right]$.
It is based on
leading order parton splitting functions and uses $Q\sim E_{\rm jet}$ as the starting
scale of the parton shower evolution. The evolution is stopped at the scale $Q_0$,
which is, like $\Lambda$, a fit parameter of order $O(\Lambda_{\rm QCD})$. 
The effects of destructive quantum interference which are responsible for the shape of the distribution translate at small $x$ into the prescription of angular ordering of a probabilistic parton cascade.

At high energies the spectrum turns out to be insensitive to the value of $Q_0$ and it is sufficient to consider the case $\lambda = 0$, which is equivalent to $Q_0 = \Lambda = Q_{\rm eff}$.
This is the so called \emph{limiting spectrum}, which for gluon jets takes the form \cite{dokshitzer:lnpi}
\begin{eqnarray}
D^{\rm lim}_g(\xi,Y) 
& = &
A\, \Gamma(B)\int^{\frac{\pi}{2}}_{-\frac{\pi}{2}} \frac{d\tau}{\pi}e^{-B\alpha}
\left[
\frac{\cosh\alpha+(1-2\zeta)\sinh\alpha}{A\, Y \frac{\alpha}{\sinh\alpha}}
\right]^{B/2} \nonumber \\
&   &
\hspace{3em} \times\, I_B
\left(
\sqrt{4\, A\, Y \frac{\alpha}{\sinh\alpha}[\cosh\alpha+(1-2\zeta)\sinh\alpha]}
\right).
	\label{eq:limspec}
\end{eqnarray}
In the MLLA approximation, the spectrum for massless quark jets differs by an overall
prefactor $C_F/N_c = 4/9$ only. Here, $I_B$ denotes the modified Bessel function of 
order $B$ and the factors $A$, $B$ are determined by the prefactors of the LO parton splitting
functions $P_{q\to qg}$, $P_{g\to gg}$ and $P_{g\to qq}$,
\begin{equation}
A = \frac{4 N_c}{b},  \hspace{2em}
B = \left(\frac{11}{3} N_c + \frac{2}{3}\frac{n_f}{N_c^2}\right)/b, \hspace{2em}
b = \frac{11}{3}N_c - \frac{2}{3}n_f,
\end{equation}
where $N_c$ is the number of colors and $n_f$ the number of flavors. In what follows we use $N_c = n_f = 3$.
We have introduced also the following notational shorthands
\begin{equation}
\label{eq:notation1}
Y = \ln\frac{E_{\rm jet}}{\Lambda}, \hspace{2em}
\alpha = \alpha_0 + i\tau, \hspace{2em}
\tanh\alpha_0 = 2\zeta -1 , \hspace{2em}
\zeta = 1 - \frac{\xi}{Y}.
\end{equation}
Local parton hadron duality (LPHD)~\cite{Dokshitzer:1991fc,Azimov:1984np,Azimov:1985by}
is a model which translates the partonic yield 
(\ref{eq:limspec}) into a hadronic one. For unidentified hadrons, LPHD postulates
a one-to-one correspondence between partons and hadrons, introducing a proportionality
factor of order $O(1)$
\begin{equation}
       \frac{dN^{\rm hadrons}}{d\xi} = 
       K_{\scriptscriptstyle \rm LPHD}\, D_{q, g}\left(\xi,Y,\lambda \right).
       \label{2.4}
\end{equation}
To describe the spectra of identified hadrons inside jets, it has been suggested to go beyond the 
limiting spectrum (\ref{eq:limspec}) by stopping the evolution of the parton shower at a finite
scale $\lambda$, related to the mass of the hadron $Q_0 \approx M_h$ \cite{Azimov:1984np, Azimov:1985by}. This leads to
 \begin{equation}
D^h(\bar{\zeta}, E_{\rm jet}, M_h) =  K_0(M_h)\, D^{\rm lim} (\bar{\zeta}, E_{\rm jet})\, ,
\label{eq:spechad}
\end{equation}
where
\begin{equation}
K_0(M_h) = 
\frac{2}{\Gamma(B)} (A\, \lambda)^{B/2}
K_B\left(\sqrt{4\, A\, \lambda}\right)\, .
\label{2.6}
\end{equation}
Here,  $K_B$ is the modified Bessel function of order $B$ and
\begin{equation}
\label{eq:zeta_y}
\bar{\zeta} = \frac{y}{y_{\rm max}}, \hspace{1cm}
y = \ln \frac{E_h+p_h}{M_h}\, .
\end{equation}
%

\begin{figure*}[t]
\centering
\includegraphics[height=0.25\textheight,width=0.46\textwidth,clip]
{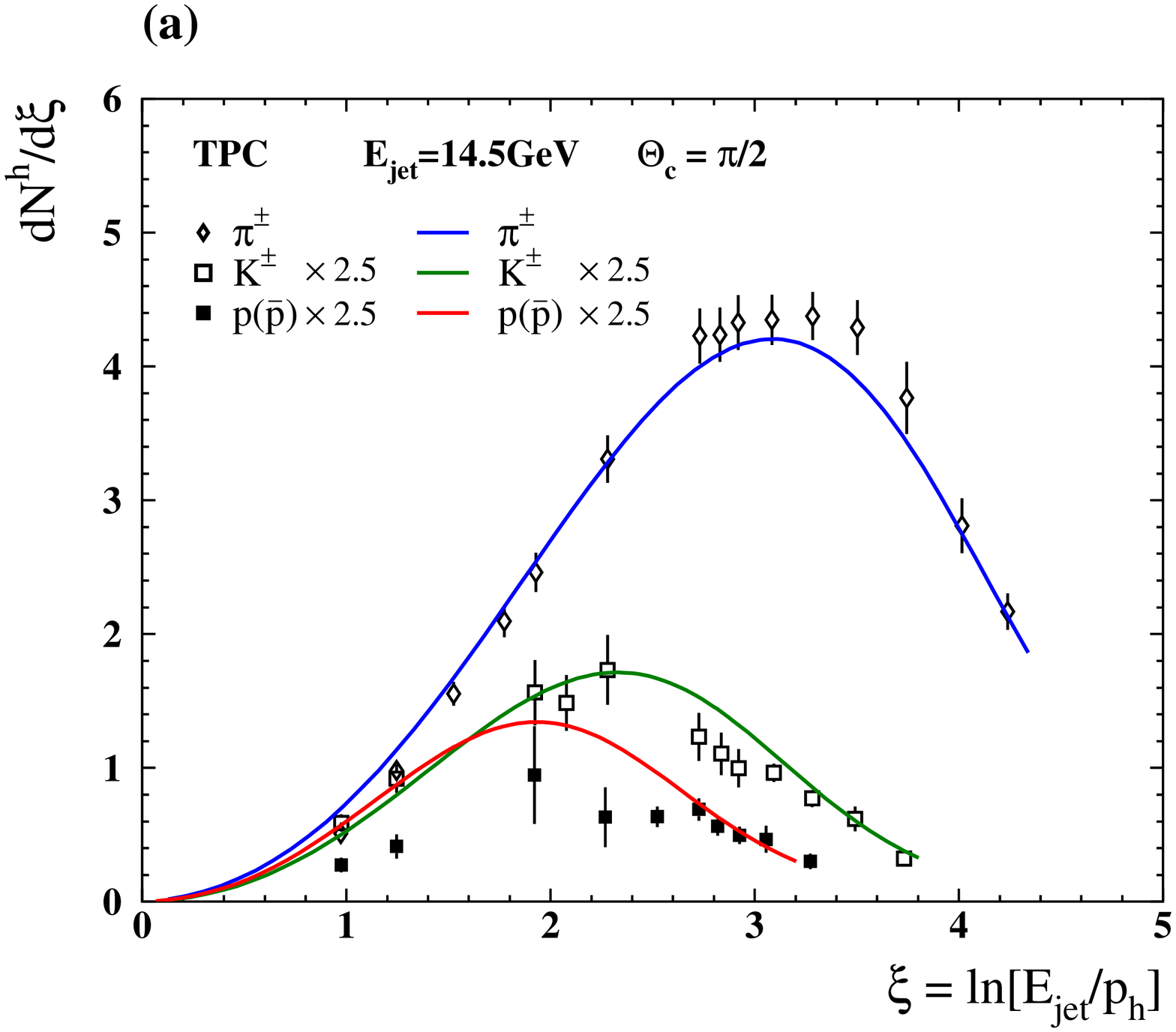}
\includegraphics[height=0.25\textheight,width=0.46\textwidth,clip]
{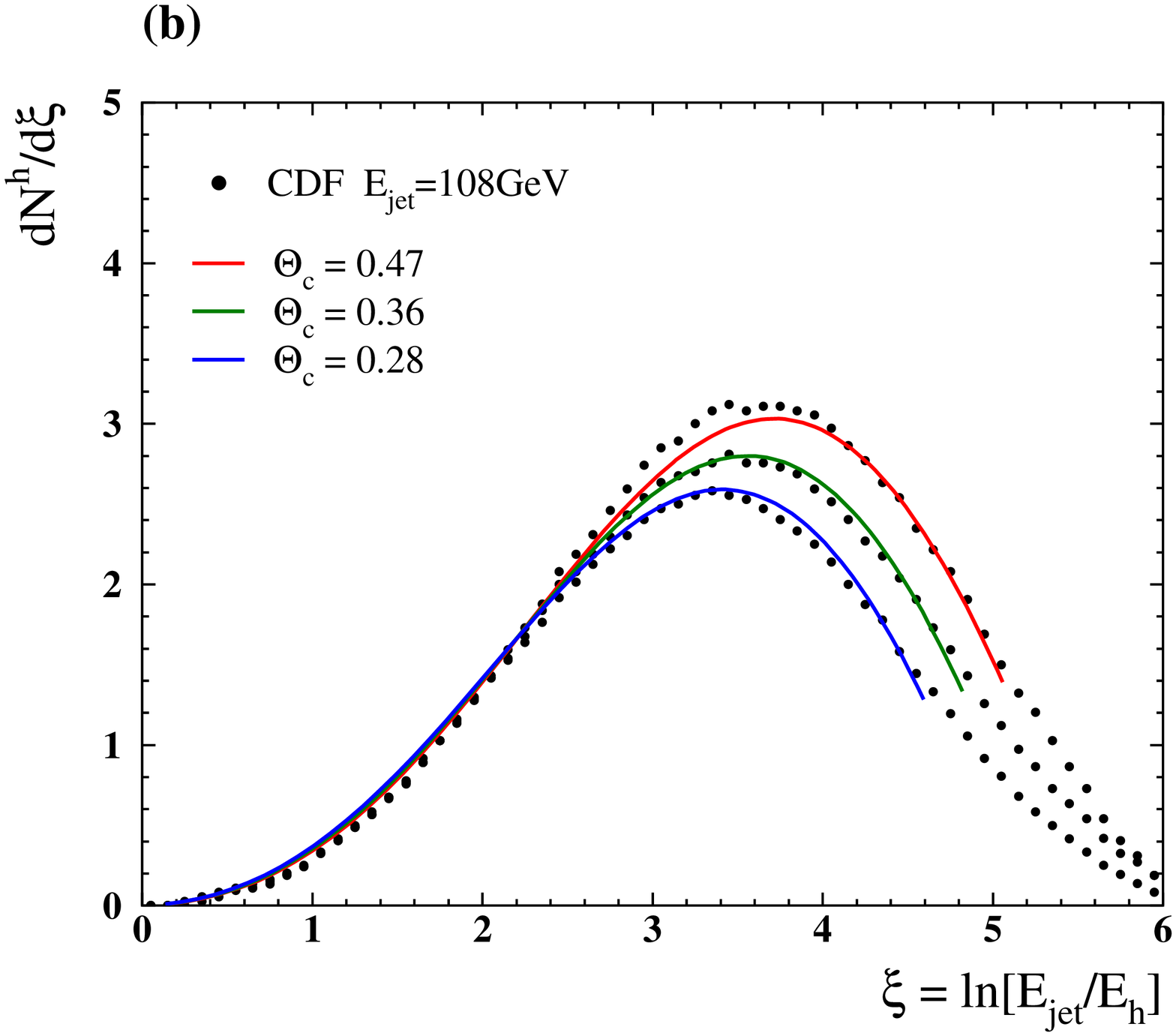}
\begin{minipage}[hbt]{6in}
\caption{Two examples for the comparison of the MLLA+LPHD formalism with data on single 
inclusive spectra inside a jet as a function of the logarithm of the hadron momentum fraction $\xi$. 
a)~The distribution  (\ref{2.4}) of charged pions ($\pi^++\pi^-$), kaons ($K^++K^-$)
and (anti)-protons ($p+\bar{p}$) in a jet of energy $E_{\rm jet} = 14.5$ GeV, compared to TPC 
data on $e^+e^-$ collisions \cite{Aihara:1983ic}. 
The MLLA parameters are $\Lambda = 155$ MeV and 
$K_{\scriptscriptstyle \rm LPHD} = 1.22$. 
b) The MLLA+LPHD distribution of all charged hadrons in a jet of energy $E_{\rm jet} = 108$ 
GeV for various opening angles $\Theta_c$, compared to CDF data from $p\bar p$ 
collisions \cite{Acosta:2002gg}. The MLLA parameters are $\Lambda = Q_0 = 235$ MeV, 
$K_{\scriptscriptstyle \rm LPHD} = 0.555$.
}
\label{fig2}
\end{minipage}
\end{figure*}

In Fig.~\ref{fig2}a, we compare  the distribution (\ref{eq:spechad}) with single inclusive 
spectra of identified hadrons, measured in jets in $e^+\, e^-$ collisions. 
We use the mass $M_h$ of the identified hadron species in the definition of the 
rapidity (\ref{eq:zeta_y}) and in the argument of the prefactor $K_0$ in (\ref{2.6}), 
Here, the limiting spectrum was calculated taking 
$Q_0 = \Lambda \approx M_{\pi} $. In accordance with data, one observes that the spectrum 
gets harder for more massive hadrons. Also, the mass-dependent hierarchy of hadron multiplicity 
is reproduced. For an improved agreement, we followed~\cite{Azimov:1985by} and multiplied
the conversion factor $K_{\rm LPHD}$ by an additional suppression factor $\gamma_s = 0.73 $
for kaons.
This heuristic factor may account for the fact that the probability of hadronizing into strange
hadrons is reduced due to the larger mass of the strange quark.
In contrast to the original analysis~\cite{Azimov:1985by}, we calculate $K_0(M_h) $ from
(\ref{2.6}) rather than extracting it from a fit to TPC data.

In Fig.~\ref{fig2}b, we compare the MLLA formalism to the single inclusive spectra of all
charged hadrons, contained inside smaller jet subcones of opening angles $\Theta_c$. 
The MLLA formalism is the result of an evolution equation which implements a
parton shower with angular ordering, and the $\Theta_c$-dependence of $D^h$ is 
given by replacing in (\ref{eq:notation1})
\begin{equation}
	E_{\rm jet} \longrightarrow E_{\rm jet} \sin\Theta_c\, .
	\label{2.3}
\end{equation}
As seen in Fig.~\ref{fig2}b, and in accordance with \cite{Acosta:2002gg}, the MLLA formalism leads to a fair 
description of the $\Theta_c$-dependence of the unidentified charged hadron spectrum 
$dN^h/d\xi$ inside a jet. 
(We note as an aside that the CDF data in Fig.~\ref{fig2}b are plotted versus the variable
$\ln[E_{\rm jet}/E_h]$. Hence, we have calculated the distributions in
Fig.~\ref{fig2}b and Fig.~\ref{fig3}a as a function of $\ln \left[E_{\rm jet}/E_h\right]$, which we 
also denote by a slight abuse of notation with the variable $\xi$. Everywhere else in
this paper, we use $\xi = \ln \left[E_{\rm jet}/p_h\right]$.)
We are not aware of an experimental study of the jet 
hadrochemical composition  as a function of the jet opening angle. In what follows, we 
assume that the relative distributions of identified hadron species inside a jet do not change 
significantly as a function of $\Theta_c$,
so that the replacements (\ref{2.3}) applies to identified yields, too.

In summary, MLLA+LPHD are known to provide a fair description of the charged and
identified single inclusive hadron spectra inside a jet, and of their dependence on jet
opening angle. This will serve as the baseline for the following study of jet
medium modifications.

\subsection{Modeling medium modifications of single inclusive intrajet distributions}
\label{sec2b}

A unique prescription of how to model the medium modification of jet
fragmentation does not exist. RHIC data indicate that a realistic prescription should lead to a softening
of the jet distribution, consistent with a suppression of the leading hadron spectra
by a factor $\sim 5$ in central Au+Au collisions. One possibility to achieve this is to 
enhance the probability of parton branching in the jet fragmentation. This is also motivated
by calculations of medium-induced gluon radiation of hard partons, which imply enhanced
parton splitting. In what follows,
we consider a model of this type \cite{Borghini:2005em}, in which the singular parts of all parton splitting
functions $P_{qq}$, $P_{gg}$ and $P_{gq}$ are enhanced by one common 
model-dependent factor $\left(1+f_{\rm med}\right)$,  such that for instance
\begin{equation}
P_{qq} = C_F\, \Bigg\{
\frac{2\, (1 + f_{\rm med })}{(1-z)_+} - (1+z)
\Bigg\}\, .
\label{2.9}
\end{equation}
The factor $\left(1+f_{\rm med}\right)$ is the only medium modification in our model. The 
LPHD-prescription will be adopted unchanged.  
%
\begin{figure*}[t]
\centering
\includegraphics[height=0.25\textheight,width=0.46\textwidth,clip]
{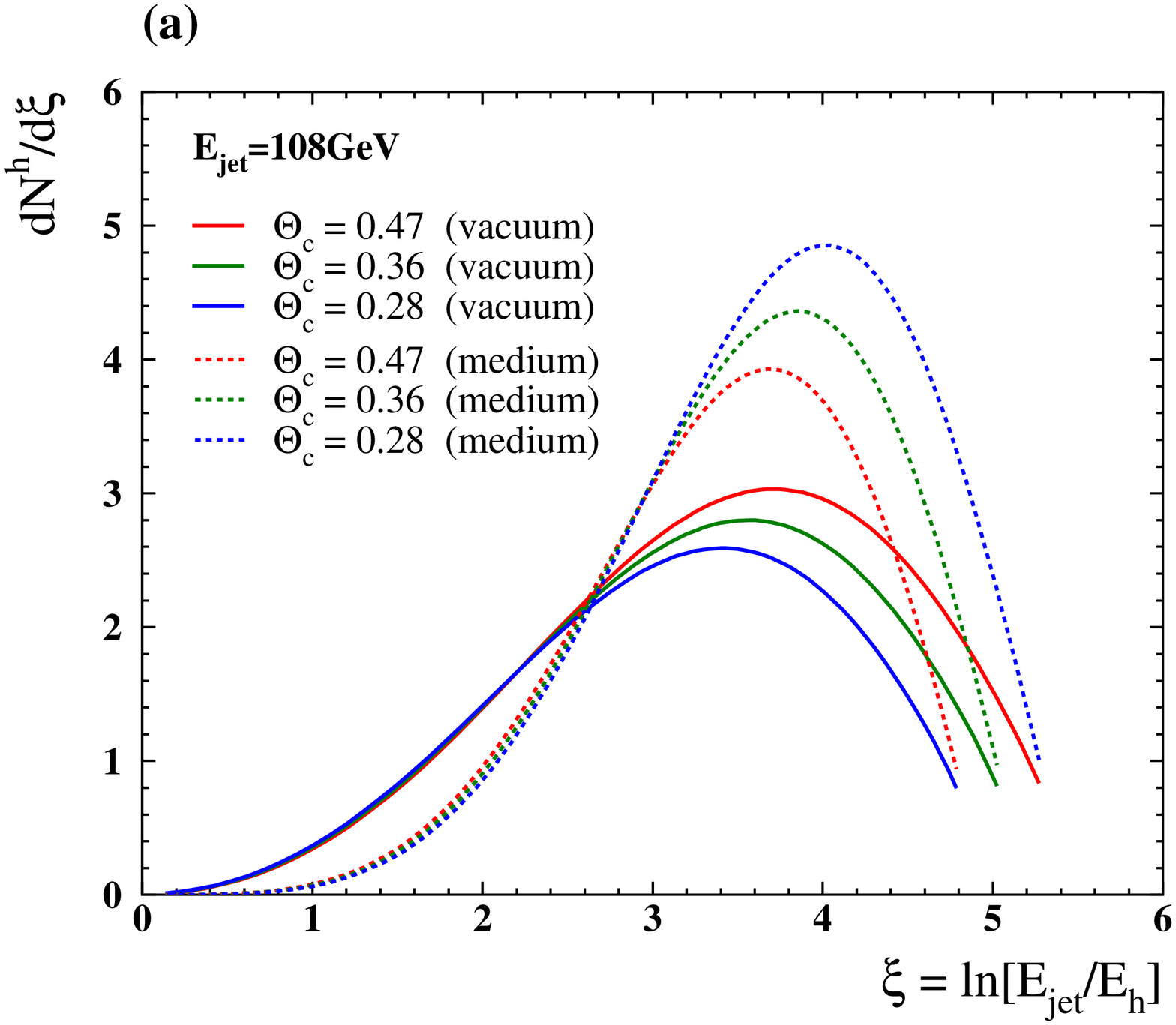}
\includegraphics[height=0.25\textheight,width=0.46\textwidth,clip]
{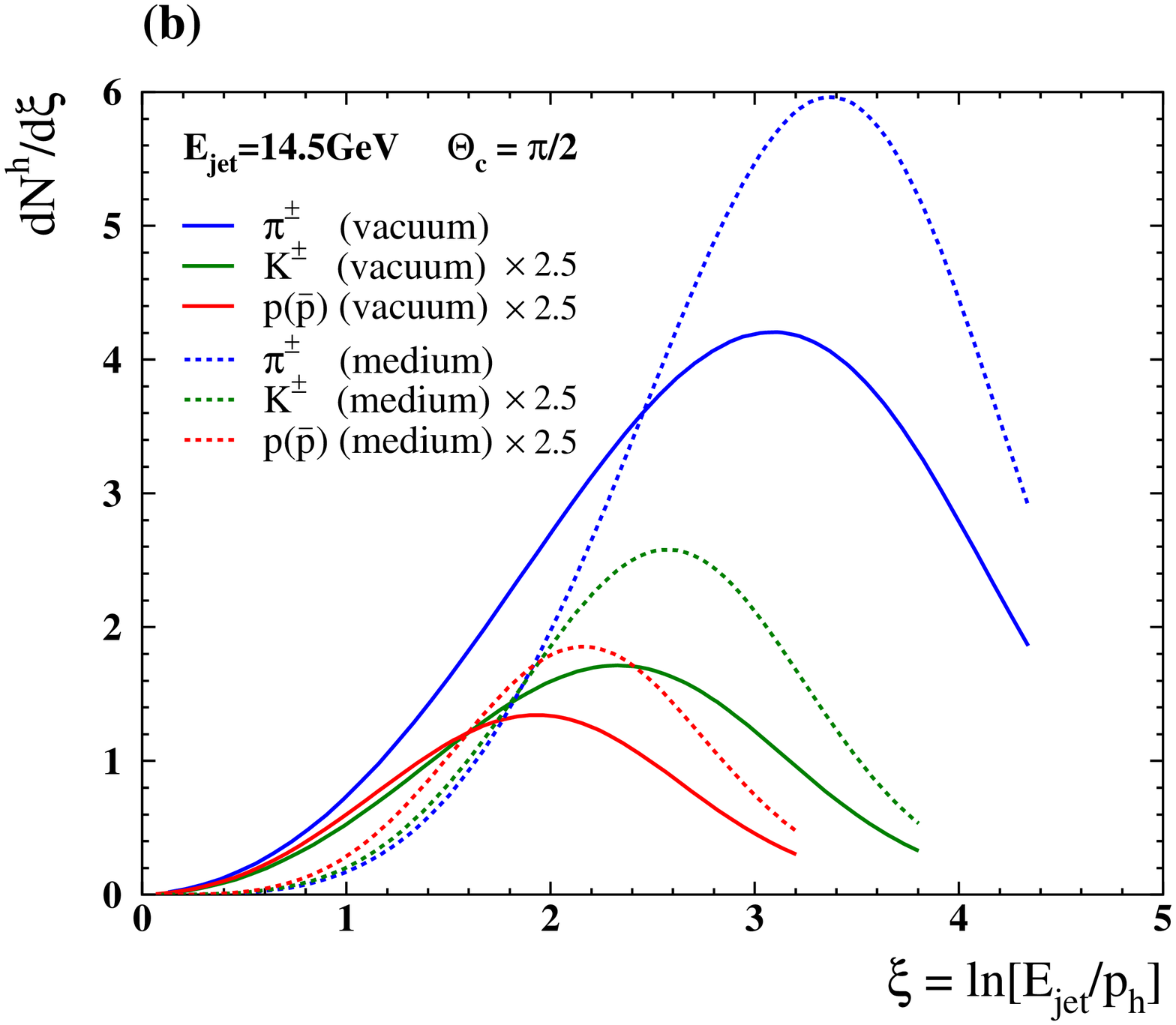}
\begin{minipage}[hbt]{6in}
\caption{Medium modification of the single inclusive spectra inside a jet as a function of the logarithm of the hadron momentum fraction $\xi$ for $f_{\rm med} = 1$.
a) Distributions of all charged particles in the jet of energy $E_{\rm jet} = 108$ GeV for various opening angles~$\Theta_c$.
b) Expected modification of the pion, kaon and proton spectra 
in the jet of energy $E_{\rm jet} = 14.5$ GeV and opening angle $\Theta_c = \pi/2$.}
\label{fig3}
\end{minipage}
\end{figure*}
%
For the present exploratory study, this model has several wanted features:
The model is easy to implement. Within the MLLA formalism, the medium-modification 
(\ref{2.9}) amounts solely to a redefinition of the parameters $A$ and $B$, 
\begin{equation}
A = \frac{4 N_c (1+f_{\rm med})}{b},  \hspace{2em}
B = 
\left(\frac{11+12\, f_{\rm med}}{3}N_c + \frac{2}{3}\frac{n_f}{N_c^2}\right)/b.
\end{equation}
The model describes not only leading fragments, but the entire jet distribution.
Also, the value of the only model parameter $f_{\rm med}$ can be constrained since it 
determines the degree to which single inclusive hadron spectra are suppressed. 
Here, we add the caveat that the MLLA spectrum becomes unreliable in the region of
large momentum fraction, $\xi < 1$ say, mainly since MLLA resums only logarithms in $1/x$ and 
not in $1/(1-x)$. The use of $D^h(\xi,Q=E_{\rm jet})$ as a fragmentation function for leading
hadron production is thus unreliable. But the finding that in this way a value 
$f_{\rm med} = 0.6-0.8$ can account for a suppression $\sim 5$ of leading hadron 
spectra, may still provide an indication of the parameter range of $f_{\rm med}$ supported
by data~\cite{Borghini:2005em}. Our only reason for mentioning this  argument is
to motivate a choice of $f_{\rm med}$. It is clear that at the LHC, experimental constraints on 
$f_{\rm med}$ will come mainly from measuring the distributions shown in Fig.~\ref{fig3},
rather than from single inclusive hadron spectra. In the absence of such constraints,
we use for the following numerical studies $f_{\rm med} =1$, which lies certainly in the 
right order of magnitude, and allows us to illustrate the features of this model.  

We now discuss the modifications of jet observables introduced by this model. 
As seen in Fig.~\ref{fig3}a, enhancing the parton splitting by a factor $\left(1+f_{\rm med}\right)$
softens the jet multiplicity distributions irrespective of the jet opening angle. Also, this
softening is reflected in all identified hadron spectra, see Fig.~\ref{fig3}b. The mass
hierarchy of the intrajet distributions is preserved in this model: the yields of heavier
hadrons peak at larger momentum fractions and thus at smaller $\xi$.
\begin{figure}[t]
\includegraphics[scale=.6,angle=0]{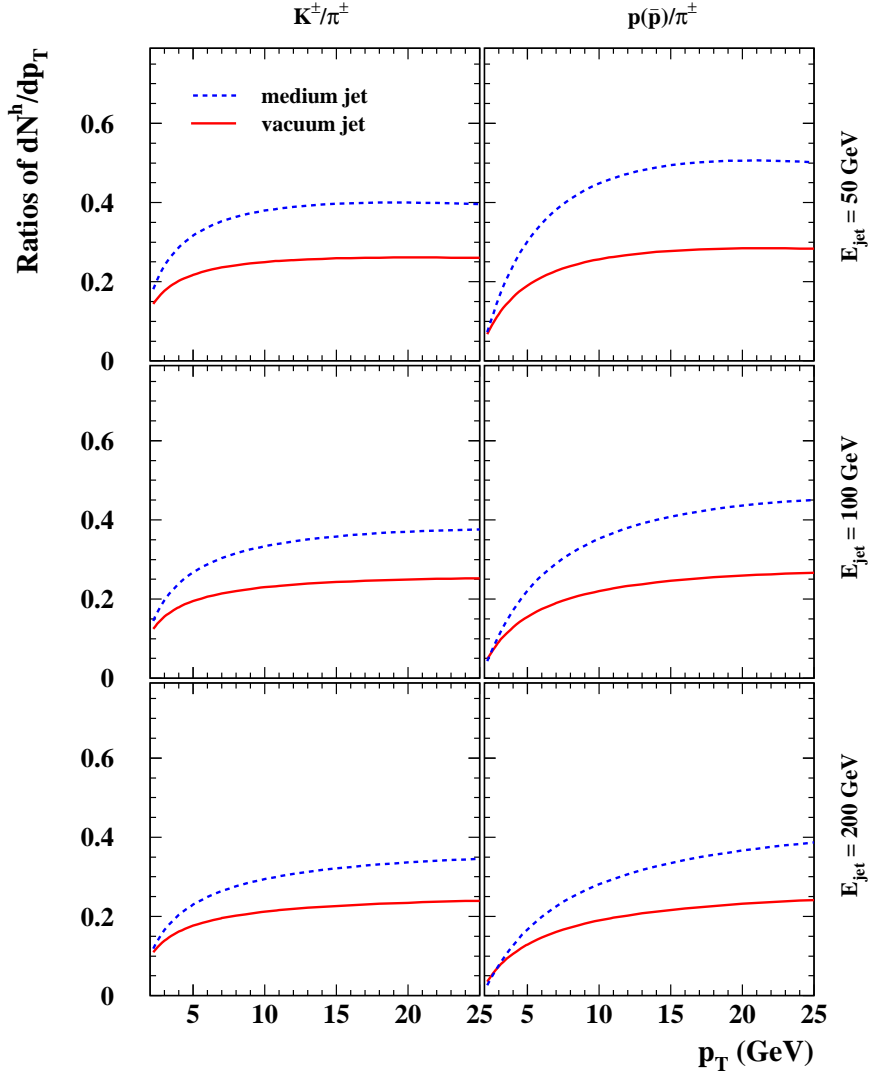}
\caption{Results of the MLLA+LPHD formalism for  $K^{\pm}/\pi^{\pm}$ and $p(\bar p)/\pi^{\pm}$ 
ratios in jets with energies $E_{\rm jet} =$ 50, 100 and 200 GeV. The jet opening angle is
$\Theta_c = 0.28$ and medium-induced changes are calculated for $f_{\rm med} = 1$.}
\label{fig4}
\vskip-0.05in
\end{figure}

To better characterize the medium modification of the jet hadrochemical composition,
implemented in this model, we focus in the following not on the absolute yields as in
Fig.~\ref{fig3}, but on the ratios of identified hadron yields. Also, we translate  
the $\xi$-dependence at fixed $E_{\rm jet}$ into a transverse momentum
dependence. For a single jet of energy $E_{\rm jet}$, the $p_T$-spectrum of identified
hadronic fragments of type $h$, collected within the opening angle $\Theta_c$, takes the
form
\begin{equation}
\Bigg[\frac{dN^h (\Theta_c)}{dp_{\scriptscriptstyle T}}\Bigg]_{\rm jet} = 
K_{\scriptscriptstyle \rm LPHD}\, \gamma_h\, K_0(M_h)\, \frac{1}{p_{\scriptscriptstyle T}} \,
D^{\rm lim}\Big(\bar{\zeta}(p_{\scriptscriptstyle T}, M_h, E_{\rm jet}),E_{\rm jet}, \Theta_c, \Lambda \Big)\, .
\label{2.11}
\end{equation}
Here,  $\gamma_h$ is an additional particle species dependent suppression factor.
We choose $\gamma_h=1$ for pions and protons, and $\gamma_K=\gamma_s=0.73$ for 
kaons~\cite{Azimov:1985by}, which are the same choices as made in section~\ref{sec2a}.
 For the local parton-hadron duality parameter, we take 
$K_{\scriptscriptstyle \rm LPHD} = 0.5$. This factor must be slightly lower than the one 
used in Fig.~\ref{fig2}b, since it determines the normalization of the identified hadron spectra,
while Fig.~\ref{fig2}b shows the spectrum of all charged particles.  
We note that for $f_{\rm med}=1$, $K_0(M_h)$ changes only by up to $\sim 12\%$ from its 
value in the vacuum.
In the following, we shall focus on results for a relatively small opening angle 
$\Theta_c = 0.28$ rad. We have tested that the dependence of identified particle ratios within the jet on $\Theta_c$ is very weak . 

In our model, the hadrochemical composition of jet fragments changes significantly 
in the presence of parton energy loss (i.e. for finite $f_{\rm med}$). Heavier hadrons
become more abundant. As seen in Fig.~\ref{fig4}, for an $E_{\rm jet} = 50$ GeV jet, the 
kaon to pion ratio 
increases by a factor $\sim 50 \%$, the proton to pion ratio by a factor $\sim 100 \%$. 
These medium-induced changes persist over the entire transverse momentum range.
They decrease slightly with increasing jet energy, but remain clearly visible even for
$E_{\rm jet} = 200$ GeV jets.

The significant medium modification of jet hadrochemistry is remarkable, since the present 
model does not encode for medium effects at or after hadronization. Also, in contrast to the 
sketch in Fig.~\ref{fig1}, the model does not involve color transfer between projectile and target, 
nor does it involve the transfer of other quantum numbers. It only encodes an enhancement 
of the probability of parton splitting, which affects the distribution of the invariant mass of partons 
at the end of the parton shower. It is one of the main results of this paper, that enhanced 
parton splitting alone without explicit medium-induced modification of the hadronization mechanism 
can lead to significant changes in the jet hadrochemical composition. In this sense, 
characteristic deviations in jet hadrochemistry are a generic characteristics
of any parton energy loss mechanism, as long as medium-modified splitting is indispensable for
parton energy loss. 

We have studied whether medium effects generally enhance the yield of heavier hadrons, as seen 
in Fig.~\ref{fig4}, or whether the opposite may be possible. In the context of MLLA+LPHD,
we observe that the yield of heavier hadrons peaks at smaller values of $\xi$ and that increasing
the splitting parameter $f_{\rm med}$ enhances the soft part of this yield. However, the
$\xi$-region which is soft for a heavy hadron is still hard for a light hadron such as a pion. 
For instance, in Fig.~\ref{fig3}b, this is the case for $1.5 < \xi < 2.5$, since the pion yield
decreases with increasing $f_{\rm med}$ for $\xi < 2.5$, while the proton yield increases
for $\xi > 1.5$. This opposite $f_{\rm med}$-dependence at intermediate
$\xi$ excludes the possibility that the ratios $K^{\pm}/\pi^{\pm}$ and $p(\bar p)/\pi^{\pm}$ 
decrease over the entire transverse momentum range with increasing medium-effects. 
This gives support to the idea that the enhancement of heavier hadrons, observed in
Fig.~\ref{fig4}, is rather generic for jet quenching models. 

\section{Medium-modified Jets in a high-multiplicity environment}
\label{sec3}

The success of MLLA+LPHD in reproducing hadrochemical distributions (see e.g. Fig.~\ref{fig2}a),
as well as the model of medium modifications described in section~\ref{sec2b}, are based on
a {\it mass effect}: Identified hadronic yields are determined by relating the QCD evolution scale
to the mass of the hadron species. In contrast, RHIC data indicate that the hadrochemical abundance 
of the underlying event in heavy ion collisions follows a {\it valence quark counting rule} in a wide
intermediate transverse momentum range: The identified hadronic yields are characterized
by a baryon-to-meson enhancement, irrespective of the hadron 
mass~\cite{Adcox:2004mh,Adams:2005dq}. So, the hadrochemical
composition of unmodified jets, and of the underlying event in heavy ion collisions appear to
differ qualitatively.

If parton energy loss constitutes the onset of a partonic equilibration mechanism, does this imply 
that the hadrochemical distribution of medium-modified jets approaches that of the medium,
rather than remaining qualitatively different? Access to these and other far reaching questions
may be gained by establishing whether the qualitative difference between jet hadrochemistry 
and bulk hadrochemistry persists at the LHC. 
Here, we shall not attempt to address these questions by building jet quenching models which implement specific features of hadrochemical equilibration. Rather, we explore in a jet quenching 
model without hadrochemical equilibration (namely that of section~\ref{sec2b}), what happens if the 
particle identified yield of a jet is superimposed to a realistic heavy ion background. This
may serve as a baseline on top of which signals of hadrochemical equilibration can be 
established. We first introduce in section~\ref{sec3a} a recombination model for the
underlying event in heavy ion collisions. Then we superimpose in section~\ref{sec3b}
jet multiplicity distributions to this underlying event.

\subsection{Two component model of underlying event}
\label{sec3a}

To model the underlying event we use the framework proposed in \cite{Fries:2003kq, Fries:2003vb} and
further explored in~\cite{Maiani:2006ia}, in which hadrons are produced via two competing mechanisms, recombination and fragmentation. Recombination models assume that hadrons
can form by coalescence of constituents quarks, which are taken as effective degrees of freedom
during hadronization. The recombination mechanism competes with the standard fragmentation 
of quarks into hadrons.
Recombination models typically discard gluonic degrees of freedom and use the same quark
spectrum for the calculation of recombination and fragmentation contributions to single
inclusive spectra. Despite these simplifications, a fair description of relative hadrochemical
yields is obtained~\cite{Fries:2003kq,Maiani:2006ia}. Here, we limit ourselves to central 
collisions and we specify the model input following Ref.~\cite{Maiani:2006ia}. 
The main input for recombination
models is knowledge about the partonic transverse momentum spectrum. This is typically
modeled by a two-component distribution, showing an exponential "thermal" slope at 
low transverse momentum and displaying a characteristic power-law at high transverse 
momentum.

We characterize the exponential component of the quark and anti-quark spectrum
by the distribution 
\begin{equation}
   w_i (R, p) \sim e^{-p^{\mu} \upsilon_{\mu}(R)/T}\, .
   \label{3.1}
\end{equation}
This distribution is assumed to be emitted from spatio-temporal positions
 $R^{\mu} = (\tau \cosh \eta, \rho \cos \phi, \rho \sin \phi, \tau \sinh \eta)$, which lie
 in a thermally equilibrated system at temperature $T$ along a space-like freeze-out
hypersurface $\Sigma$. Here, $\eta$ is the space-time rapidity, $\rho$ the radial coordinate,
and a suitable hypersurface can be specified by fixing  $\tau = \sqrt{t^2 - z^2}={\rm const}$.
The system is expanding at position $R^\mu$ with a longitudinally boost-invariant
flow profile $\upsilon_{\mu}(R)$, which displays a velocity $\upsilon_{\scriptscriptstyle T} = \tanh \eta_{\scriptscriptstyle T}$ in the transverse direction.
Integrating $w_i$ along the freeze-out hypersurface, it is a standard procedure to find
the soft contribution to the quark spectrum, $d N_a^{\rm soft}/d^2 p_{a,\scriptscriptstyle T}\, dy$.
In the following, we denote by $p_{a,\scriptscriptstyle T}$ the momentum of partons, and 
by $p_T$ the momentum of hadrons.

The hard, power-law contribution to the quark spectrum is determined by
\begin{equation}
  \frac{d N_a^{\rm hard}}{d^2 p_{a,\scriptscriptstyle T}\, dy} \Bigg|_{y=0} =
  K \frac{C}{(1+p_{a,\scriptscriptstyle T}/B)^\beta}.
  \label{3.2}
\end{equation}
Here, the parameters $C$, $B$ and $\beta$ are taken from leading order perturbative QCD 
calculations \cite{Srivastava:2002ic} and the constant $K\simeq 1.5$ accounts for higher order 
corrections~\cite{Fries:2003kq, Fries:2003vb}. Parton energy loss is modeled by quenching this
partonic spectrum via shifting its momentum distribution by 
$\Delta p_{a,\scriptscriptstyle T}(p_{a,\scriptscriptstyle T}) = \epsilon_0 \,\sqrt{p_{a,\scriptscriptstyle T}}$, as suggested in~\cite{Baier:2001yt}.

We now explain how these partonic spectra are turned into hadronic yields. For an
exponential spectrum based on (\ref{3.1}), recombination always wins over fragmentation,
since there are exponentially many recombination partners at soft $p_{T}$. For a power-law
tail (\ref{3.2}), however, fragmentation wins over recombination, since there are sufficiently
many high-$p_{T}$ components which can fragment into softer ones. Thus, the partonic 
$p_{a, T}$-scale at which the power-law contribution (\ref{3.2}) overcomes the exponential 
one sets the hadronic $p_{T}$-scale at which fragmentation starts to dominate over
recombination~\cite{Fries:2003kq}.

The momentum spectrum for mesons and baryons from recombination can be written 
as~\cite{Fries:2003kq,Maiani:2006ia}
\begin{equation}
\label{eq:backspec1}
\frac{dN_{M,B}}{d^2 p_{\scriptscriptstyle T}\, dy}\Bigg|_{y=0} =   
C_{M,B} M_{\scriptscriptstyle T} 
\frac{\tau A_{\scriptscriptstyle T}}{(2\pi)^3} \, 2\, \Pi_a \, \gamma_a \,   
I_0 \left[ \frac{p_{\scriptscriptstyle T}\sinh \eta_{\scriptscriptstyle T}}{T}\right]   k_{2, 3}(p_{\scriptscriptstyle T}) \, ,   
\end{equation} 
where $\gamma_a$ are quark fugacities, $C_{M,B}$ the degeneracy factors for mesons and baryons respectively, and $M_{T}$ their transverse masses. $A_{\scriptscriptstyle T} = \pi \rho^2_0$ is the transverse area 
of the parton system at freeze-out and $\tau$ the hadronization time. Here, we introduced the 
shorthand $k_N(p_{\scriptscriptstyle T}) =
  K_1 \left[ \frac{\cosh \eta_{\scriptscriptstyle T}}{T}\sum_{a=1}^N\sqrt{m_a^2 + \frac{p_{\scriptscriptstyle T}^2}{N^2}}\right]$.

The  spectrum for hadrons from fragmentation is given by
\begin{equation}
\label{eq:hadfrag}
  E \frac{d N_h}{d^3 p_{\scriptscriptstyle T}} = \sum_a \int\limits_0^1 \frac{d z}{z^2}
  D_{a\to h}(z, Q^2) E_a \frac{d N_a^{\rm hard}}{d^3 p_{a, \scriptscriptstyle T}}
\end{equation}
with $D_{a\to h}(z, Q^2)$ denoting the fragmentation function of a parton $a$ into  a hadron $h$. 
We use KKP fragmentation functions~\cite{Kniehl:2000fe}.

It has been shown~\cite{Fries:2003kq, Fries:2003vb,Maiani:2006ia} that with appropriately chosen parameters,
this two component model accounts successfully for the baryon-to-meson enhancement 
observed in a large class of RHIC data on Au+Au collisions at intermediate $p_{\scriptscriptstyle T}$.
In particular, recombination models can reproduce the proton to pion and kaon to pion ratio
at intermediate transverse momentum~\cite{Fries:2003vb,Fries:2003kq, Maiani:2006ia}. Recombination 
dominates at RHIC up to $p_{\scriptscriptstyle T}^{\rm hadron} \simeq 4 - 6$ GeV, and 
fragmentation takes over for higher transverse momentum.

This model has been extrapolated to Pb+Pb collisions at 
$\sqrt{s} = 5.5\ {\rm TeV}$~\cite{Fries:2003fr, Maiani:2006ia} by fixing  the temperature of the quark phase at hadronization at 175 MeV, similar to the RHIC case, and rescaling the parameters 
$\upsilon_T$ and $\tau A_T$ such that the results of fluid simulations \cite{Eskola:2005ue} are reproduced:
$\upsilon_{\scriptscriptstyle T} = 0.68$ and $\tau A_{\scriptscriptstyle T} = 11.5 \times 10^3$ fm$^3$ \cite{Maiani:2006ia}.
The quenching of high-$p_{a,{\scriptscriptstyle T}}$ partons is fixed by the choice $\epsilon_0=2.5$, which amounts
to a factor $\simeq 10$ suppression of the single inclusive hadron spectra at 
$p_{{\scriptscriptstyle T}}=10$ GeV.
Our description of the hadrochemical composition of the underlying event in heavy ion collisions
is based on this model. The single inclusive hadron spectra calculated such for LHC are 
dominated by recombination up to a scale which lies $\simeq 2$ GeV higher than the corresponding
scale at RHIC \cite{Fries:2003fr, Maiani:2006ia}.
\begin{figure}[t]
\includegraphics[scale=.6,angle=0]{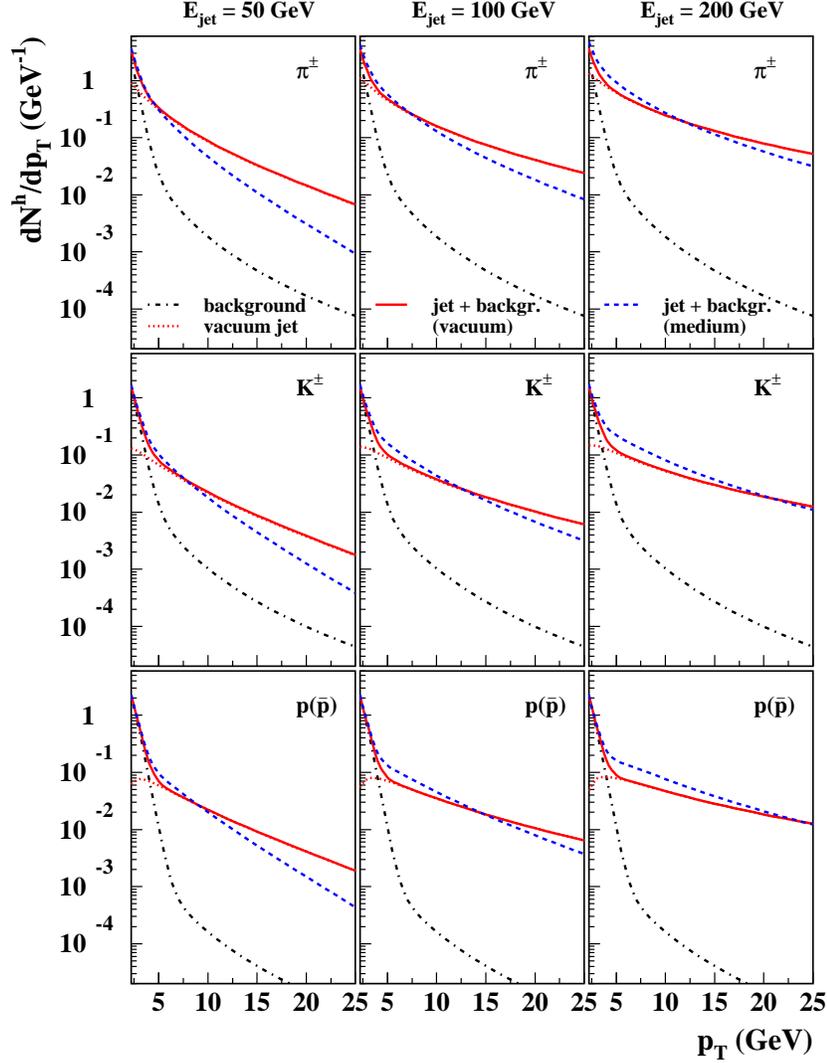}
\caption{Identified transverse momentum spectra within a cone of opening angle $\Theta_c=0.28$
for pions, kaons and protons.}
\label{fig5}
\vskip-0.05in
\end{figure}

\subsection{The hadrochemical composition of jets within high-multiplicity nucleus-nucleus collisions}
\label{sec3b}

In this section, we superimpose the jet spectrum calculated in section~\ref{sec2b} onto 
the background spectrum introduced in section~\ref{sec3a}. 
The background yield in a cone of opening angle $\Theta_c$ is given by
\begin{equation}
\Bigg[\frac{dN^h (\Theta_c)}{dp_{\scriptscriptstyle T}}\Bigg ]_{\rm background}  
\simeq \
\Theta_c^2\, \pi\, p_{\scriptscriptstyle T}\, \frac{dN^h}{d^2 p_{\scriptscriptstyle T}\, dy}\Bigg|_{y=0}.    
\label{3.5}
\end{equation}
where the spectrum on the right hand side is a sum of contributions from recombination (\ref{eq:backspec1}) and fragmentation (\ref{eq:hadfrag}).
To arrive at this expression, we have integrated the full double-differential spectrum
$ dN^h/d^2 p_{\scriptscriptstyle T}\, dy |_{y=0}$ over one unit in rapidity
and the full azimuthal phase space. Since the spectrum
is flat around mid-rapidity, the integral is trivial.
In the $\Delta\eta \times \Delta\phi$-space, this is an
area of $2\pi$. We have then multiplied our result with the fraction $\pi\Theta_c^2 / 2\pi$
which a cone of opening angle $\Theta_c$ occupies in this plane. 

A jet of energy $E_{\rm jet}$ will deposit within the opening angle $\Theta_c$ the
particle yield $\frac{dN^h (\Theta_c)}{dp_{\scriptscriptstyle T}}$ given in Eq. (\ref{2.11}). On average, this
jet will sit on top of the background (\ref{3.5}). In Fig.~\ref{fig5}, we compare these
two contributions. One sees that despite the high multiplicity environment of a heavy 
ion collision, the harder distribution of jet fragments dominates rapidly over the 
abundant distribution of soft particles at transverse momenta larger than $5-7$~GeV.
In the high-$p_T$ region, the slope of the combined transverse momentum spectrum
is dominated by jet fragments. This slope steepens characteristically in the presence 
of medium-induced parton energy loss. Hence, if the energy of the jet can be measured reliably in
heavy ion collisions, then such transverse momentum spectra provide direct 
experimental access to the longitudinal jet fragmentation function. 

\begin{figure}[t]
\includegraphics[scale=.6,angle=0]{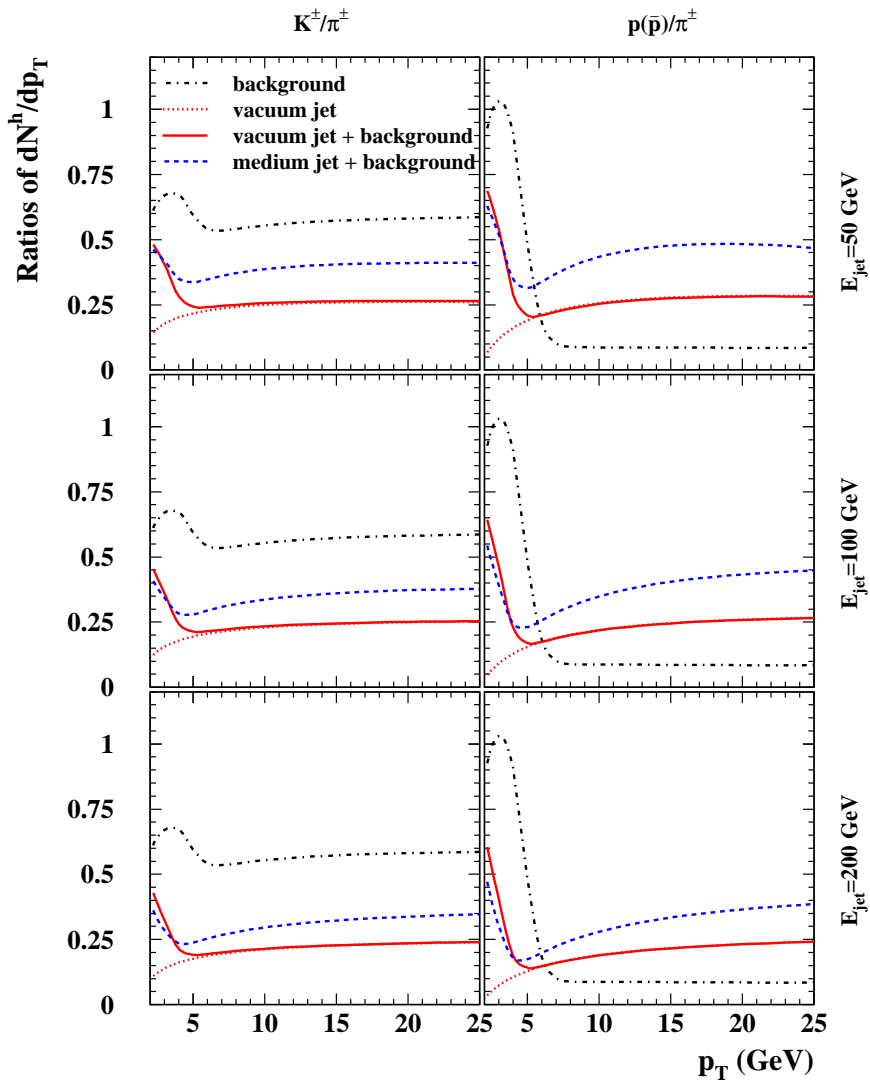}
\caption{The particle ratios $K^{\pm}/\pi^{\pm}$ and $p(\bar p)/\pi^{\pm}$ obtained
from the spectra shown in Fig.~\ref{fig5}. These ratios are measured in a cone of
opening angle $\Theta_c=0.28$ in the $\Delta\eta \times \Delta\phi$-plane, which contains both
soft background and a jet of energy $E_{\rm jet}$.}
\label{fig6}
\vskip-0.05in
\end{figure}

In Fig.~\ref{fig6}, we have plotted the identified particle ratios $K^{\pm}/\pi^{\pm}$ and 
$p(\bar p)/\pi^{\pm}$, measured in a cone of opening angle $\Theta_c$. From an
experimental point of view, this may be a relatively straightforward hadrochemical 
measurement: It is performed by counting all particles within an opening angle $\Theta_c$
as a function of $p_T$. Particle identification is statistical. An experimental separation
of the jet from the underlying event is not needed. In principle, the measurement
requires knowledge of the energy of the jet contained in the opening angle. However,
this determination of $E_{\rm jet}$ does not need to be particularly good, since the
hadronic ratios shown in Fig.~\ref{fig6} depend only weakly on $E_{\rm jet}$.

We remarked in our discussion of Fig.~\ref{fig5} that in the presence of a jet, the
spectrum within an opening cone is essentially background free above $p_T > 5-7$ GeV.
In Fig.~\ref{fig6}, this is reflected in the fact that above this transverse momentum scale, 
the particle ratios match
those shown in Fig.~\ref{fig4} and calculated without including the background. It is 
remarkable that in this high-$p_T$ range, medium-effects enhance these ratios, thus
widening rather than narrowing the difference with the background. At lower transverse
momentum, the background yield dominates the hadronic abundances and particle ratios.

We finally explore yet another possibility of presenting information about the
hadrochemical composition in jets and their change in a medium. To this end,
we introduce, in close analogy with the nuclear modification factor, the 
jet modification factor
\begin{equation}
J_{\rm AA} \equiv  
\frac{\frac{dN^h}{dp_{\scriptscriptstyle T}}\Big | _{\rm med}}{\frac{dN^h}{dp_{\scriptscriptstyle T}}\Big | _{\rm vac}}\, .
\label{3.6}
\end{equation}
Here, the numerator denotes the particle spectrum within a cone of opening angle $\Theta_c$
in the presence of a jet. We determine it as the sum of the background (\ref{3.5}) and the
spectrum (\ref{2.11}) for a quenched jet ($f_{\rm med} = 1$). The denominator is constructed
experimentally by measuring the minimum bias spectrum in a cone of angle $\Theta_c$ and 
adding the spectrum of an unquenched jet. In our calculation, we add on top of the background 
(\ref{3.5}) the spectrum (\ref{2.11}) for a vacuum jet ($f_{\rm med} = 0$).

\begin{figure}[t]
\includegraphics[scale=.59,angle=0]{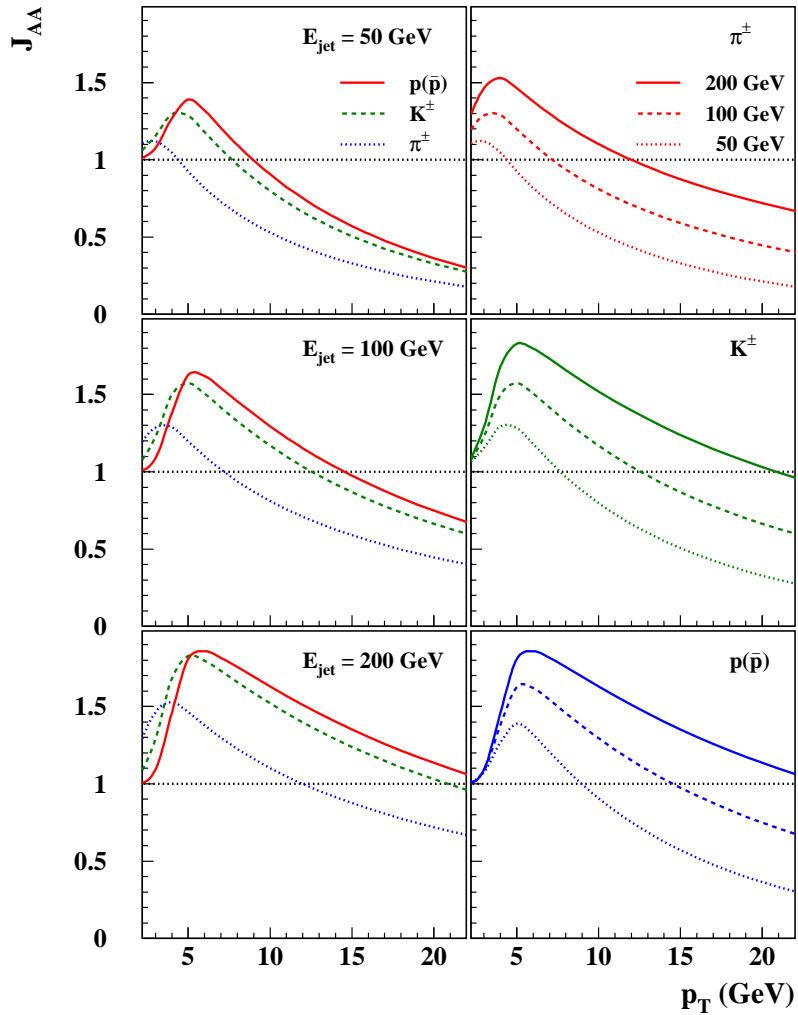}
\caption{The jet modification factor $J_{AA}$, defined in (\ref{3.6}) within a cone of opening angle $\Theta_c=0.28$ as a function of transverse 
momentum for different jet energies and different hadron species. }
\label{fig7}
\vskip-0.05in
\end{figure}

Jet quenching amounts to a reshuffling of hadronic yield from high to low transverse 
momentum. The transverse momentum scale $p_T^{\rm crit}$ above which the yield in 
a jet spectrum is depleted is characterized in the measurement of $J_{\rm AA}$ by 
 $J_{\rm AA}(p_T^{\rm crit})=1$. From the right column of Fig.~\ref{fig7}, one sees that 
 $p_T^{\rm crit}$ 
 grows significantly with $E_{\rm jet}$. This indicates that the subleading jet fragments,
 additionally produced due to medium effects, are distributed in an increasingly wide
 transverse momentum regime which extends significantly beyond the low-$p_T$
 region dominated by background yield, and up to $p_T^{\rm crit}$. Also, the total
 amount of additional multiplicity, produced due to parton energy loss, increases with
 $E_{\rm jet}$, and so  $J_{\rm AA}$ increases with $E_{\rm jet}$ in a wide intermediate
 transverse momentum regime in which hadrochemical identification is possible. 
 The order of the particle species dependence of $J_{\rm AA}$, seen in the left
 column of Fig.~\ref{fig7}, is a direct consequence of the medium-induced enhancement of
 the ratios $K^\pm/\pi^\pm$ and $p\, (\bar{p})/\pi^\pm$, seen in Figs.~\ref{fig5} and \ref{fig6}.
 Namely, if for instance at fixed transverse momentum, 
$\langle K^\pm\rangle_{\rm med}  / \langle \pi^\pm \rangle_{\rm med} > 
\langle K^\pm\rangle_{\rm vac}  / \langle \pi^\pm \rangle_{\rm vac}$,
then
$\langle K^\pm\rangle_{\rm med}  /  \langle K^\pm\rangle_{\rm vac}    > 
 \langle \pi^\pm \rangle_{\rm med}  / \langle \pi^\pm \rangle_{\rm vac}$,
 and this order is reflected in Fig.~\ref{fig7}.

\section{Summary}
\label{sec4}

The analysis of the hadrochemical composition of jets in heavy ion collisions is a \emph{terra
incognita}, for which data are not available yet. It is in experimental reach of the LHC.
In this paper, we have argued that jet quenching generically implies modifications of the jet
hadrochemical composition (see discussion in sections~\ref{sec1} and \ref{sec2b}). A
simple model of a parton shower with enhanced medium-induced branching but 
unmodified hadronization (see section~\ref{sec2b}) then allowed us to illustrate that a
significantly modified jet hadrochemistry can be expected even if at the time of
hadronization, medium effects are not present any more. This is important, since 
formation time arguments indicate that for sufficiently large $E_{\rm jet}$, hadronization
occurs outside the medium produced in heavy ion collisions.

Remarkably, at sufficiently high $p_T$, we observed that medium modifications
increase the hadrochemical ratios $K^{\pm}/\pi^{\pm}$ and $p(\bar p)/\pi^{\pm}$,
thus further increasing the difference between the distributions inside the jet
and those of the underlying event. We gave arguments of why this is likely to be
a generic feature, which should persist for a wide class of jet quenching models.
On the other hand, any jet quenching mechanism, which for instance kicks components
of the background into the jet cone, may be expected to have the opposite effect,
namely to narrow the difference between the hadrochemical composition of the 
jet and the background. It is in this sense, that we view the model study presented
in sections~\ref{sec2b} and \ref{sec3} as a baseline, on top of which effects indicative of specific
microscopic mechanisms of parton energy loss may be established.

\begin{acknowledgments}
We have profited from discussions with Nicolas Borghini, Yuri Dokshitzer, 
Krzysztof Golec-Biernat, Andreas Morsch, Karel Safarik, J\"urgen Schukraft, 
and Yves Schutz. We thank Steffen Bass, Rainer Fries and Carlos Salgado
for sharing with us the results of simulations, documented in 
Refs.~\cite{Fries:2003kq,Maiani:2006ia}. We acknowledge
support from the Marie Curie ESRT Fellowship of the European Community's
Sixth Framework Programme under contract number (MEST-CT-2005-020238),
which made this project at CERN possible. One of us (S.S.) also acknowledges
a grant of the Polish Ministry of Science, N202 048 31/2647 (2006-08).
\end{acknowledgments}


\end{document}